# Pressure-driven Superconductivity in Transition-metal Pentatelluride HfTe$_5$

Yanpeng Qi[1], Wujun Shi[1,2], Pavel G. Naumov[1], Nitesh Kumar[1], Walter Schnelle[1], Oleg Barkalov[1], Chandra Shekhar[1], Horst Borrmann[1], Claudia Felser[1], Binghai Yan[1,2,3]*, Sergey A. Medvedev[1]*

[1]Max Planck Institute for Chemical Physics of Solids, 01187 Dresden, Germany.
[2]School of Physical Science and Technology, ShanghaiTech University, Shanghai 200031, China.
[3]Max Planck Institute for the Physics of Complex Systems, 01187 Dresden, Germany.

**Abstract**

**Layered transition-metal tellurides have attracted considerable attention because of their rich physics; for example, tungsten ditelluride WTe$_2$ exhibits extremely large magnetoresistance; the tritelluride ZrTe$_3$ shows a charge density wave transition at low temperature; and the pentatelluride ZrTe$_5$ displays an enigmatic resistivity anomaly and large thermoelectric power. Recently some transition-metal tellurides are predicted to be quantum spin Hall insulators (e.g. ZrTe$_5$ and HfTe$_5$) or Weyl semimetals (e.g. WTe$_2$ and MoTe$_2$) and were subjected to intensive investigations. Here, we report on the discovery of superconductivity in hafnium pentatelluride HfTe$_5$ under high pressure. We observe two structural phase transitions and metallization with superconductivity developing at around 5 GPa. A maximal critical temperature of 4.8 K is attained at a pressure of 20 GPa, and superconductivity persists up to the maximum pressure in our study (42 GPa). Theoretical calculations indicate that the superconductivity develops mainly in the Te atom layers at medium pressure and in the Te atom chains at high pressure.**

---

* E-mail: Sergiy.Medvediev@cpfs.mpg.de; Yan@cpfs.mpg.de

The layered transition-metal tellurides, MTe$_n$ (M is a transition-metal, $n$ = 2, 3, 5), have been intensively studied for their rich physics. MTe$_n$ represents many compounds by combination of the elements (M) and stoichiometry ($n$). Ditellurides, such as WTe$_2$ and MoTe$_2$, exhibit extremely large magnetoresistance [1,2] and are predicted to be Weyl semimetals[3,4] and quantum spin Hall insulators[5] in bulk and monolayer form, respectively, which have promising potential applications in electronic, and spintronics[6,7]. Tritellurides, typified by ZrTe$_3$, present a Peierls instability and a charge density wave (CDW) transition at low temperature[8]. More interesting, bulk superconductivity can be observed upon suppression of CDW order in ZrTe$_3$[9]. Pentatellurides MTe$_5$ (M = Zr or Hf) are the highest tellurides in MTe$_n$. Particularly, MTe$_5$ have been previously investigated for the enigmatic resistivity anomaly[10,11,12], thermoelectric properties[13] and quantum oscillations[14,15]. Recently, ab-initio calculations indicated that single–layer MTe$_5$ is a large-gap quantum spin Hall insulator[16]. However, more recent angle-resolved photoemission spectroscopy experiments speculated ZrTe$_5$ to be a 3D Dirac semimetal[17]. Moreover, a chiral magnetic effect associated with the transformation from a Dirac semimetal to a Weyl semimetal was observed on ZrTe$_5$ in magneto-transport measurement[17,18]. MTe$_5$ compounds were predicted to locate close to the phase boundary between the weak and strong topological insulators and supply a platform to study topological quantum-phase transitions[16]. As a powerful tool to tune the electronic properties, pressure was also applied to these compounds. Recently, a pressure-induced semimetal to superconductor transition was observed in ZrTe$_5$[19], while the behavior of HfTe$_5$ under pressure is yet to be explored.

In this work, we address the above mentioned issues by investigating the high-pressure behavior of hafnium pentatelluride HfTe$_5$. Through electrical transport and Raman scattering measurements, we find superconductivity in two high-pressure phases of HfTe$_5$ with different normal-state features. Often superconductivity emerges in transition-metal chalcogenides when a resistivity maximum or CDW transition is suppressed by applied pressure. This seems to be also the case for HfTe$_5$. Superconductivity emerges at a pressure of 5 GPa, exhibits a maximal critical temperature ($T_c$) of 4.8 K at 20 GPa, and persists till the highest measured pressure of 42 GPa. Our results indicate that high-pressure studies offer a unique opportunity for uncovering novel physical properties in topological bulk materials.

**Results**

**Sample characterization at ambient pressure.** Single crystals of HfTe$_5$ were grown by the Te flux method. Energy-dispersive X-ray analysis confirms that the single crystals are homogeneous and the atomic ratio of elements is Hf:Te = 1 : 4.97(2). Single crystal X-ray diffraction demonstrates that our HfTe$_5$ samples adopt the *Cmcm* structure with lattice parameters *a* = 3.974(1) Å, *b* = 14.481(2) Å, *c* = 13.720(2) Å, in good agreement with previously reported structural data[20]. The crystal structure of HfTe$_5$ is shown in Figure 1a, b. The HfTe$_3$ prisms and the zigzag chains are connected through the apical Te atoms, and the Te-Te bond length between two chains is longer than that in the zigzag chain. Each HfTe$_5$ layer is nominally charge neutral, and the interlayer distance (along the *b* axis) is quite large (about 7.24 Å), suggesting a weak interlayer coupling, presumably of van der Waals type.

The specific heat, $C_P(T)$, of HfTe$_5$ at low temperatures (Supplementary Figure 1) does not follow the Debye $T^3$ law for 3D crystals. Instead, in the temperature range 1.9-6.9 K a power law $C_P(T) \propto T^\alpha$ with $\alpha$ = 2.7 is observed. Thus, in agreement with the crystal structure and as already concluded in earlier investigations[21], HfTe$_5$ displays a quasi-2D anisotropy. No contribution linear in *T* to $C_P(T)$ is observed, indicating the absence of conduction electrons in HfTe$_5$ at ambient pressure.

Figure 1c shows the temperature dependence of the electrical resistivity in the chain direction for a HfTe$_5$ single crystal at ambient pressure. The curve shows a pronounced anomalous peak near 40 K, in agreement with published data[11, 12, 22]. A similar resistive anomaly is observed in related zirconium pentatelluride ZrTe$_5$[10, 11]. This anomaly in pentatellurides is likely associated with peculiarities of their electronic structure although the origin still remains elusive[23, 24, 25]. Similarities in structure between the pentatellurides and chalcogenides strongly suggested the formation of a CDW as the origin of the resistivity anomalies. However, a search for direct evidence of CDWs in pentatellurides failed to demonstrate them[24]. Very recently, Zhao *et al.* reported that a 3D topological Dirac semimetal state emerges at temperatures around the resistivity peak, which they considered to indicate the topological quantum phase transition between two distinct weak and strong TI phases in HfTe$_5$[26]. By approaching the topological critical point, the bulk band gap goes to zero, thereby giving rise to a pronounced resistivity peak. Our HfTe$_5$ crystals, in addition, display a quite large unsaturated magnetoresistance of 5100 % at 2 K in a magnetic field of 9 T as shown in Supplementary Figure 2. A large anisotropy in the electrical transport is also observed which is typical for layered transition-metal chalcogenides[27].

**Pressure-induced superconductivity.** The resistivity anomaly in HfTe$_5$, even if its nature remains elusive, indicates that HfTe$_5$ is located in the vicinity of an electronic instability. It is well known, that superconductivity often appears in compounds which are close to a structural, magnetic, or electronic instability. In this respect, pressure can effectively modify lattice structures and influence the corresponding electronic states in a systematic fashion. Hence, the electronic transport of HfTe$_5$ has been studied as function of temperature at different pressures ($P$). Figure 2a shows the evolution of temperature dependence of electrical resistivity $\rho(T)$ for pressures up to 42 GPa. For $P <$ 5 GPa, $\rho(T)$ displays a semiconducting-like behavior similar to that observed at ambient pressure, albeit with a broadened and less pronounced anomaly. With increasing $P$, the temperature of resistivity peak increases to ~110 K at $P \approx$ 5.0 GPa, but then shifts back towards lower temperatures at further $P$ increase. A similar pressure dependence of the resistivity anomaly is observed for ZrTe$_5$[19]. This obviously common pressure behavior of the two pentatellurides is different from that of chalcogenides exhibiting a CDW (CDW transitions are typically suppressed by application of pressure)[28]. Surprisingly, the onset of superconductivity is observed at $T_c$ = 1.8 K as $P$ increases above 5 GPa. At this pressure the normal state still exhibits a pronounced resistivity anomaly at ≈ 90 K as seen inset of Figure 2a. With further increasing $P$ the resistivity anomaly is suppressed further and for $P >$ 9 GPa the temperature dependence of $\rho(T)$ changes to that of a normal metal. The critical temperature of superconductivity, $T_c$, gradually increases with $P$ and the maximum $T_c$ of 4.8 K is attained at $P \approx$ 20 GPa as shown in Figure 2b. Beyond this $P$, $T_c$ decreases very slowly and persists with $T_c$ = 4.5 K up to the highest attained pressure of 42 GPa. The pressure evolution of $T_c$ in HfTe$_5$ is very similar to that of the superconducting phase I (SC-I) of ZrTe$_5$[19]. In contrast, we do not observe any indication of a second superconducting phase analogous to the SC-II phase in ZrTe$_5$[19]. Considering the close similarities in ambient-pressure properties and the pressure-driven behavior of both pentatellurides, it might be supposed that the SC-II phase in ZrTe$_2$ is rather a metastable state characteristic only for this compound. The negligible variation of $T_c$ over a very large range of pressure observed for both pentatellurides is highly unusual, however a similar effect was observed for the pressure-induced superconductivity in some topologically nontrivial systems such as $Bi_2Se_3$[29] and $BiTeCl$[30].

The appearance of superconductivity in $HfTe_5$ is further corroborated by the resistivity data in applied magnetic fields. As seen from Figure 2c, the superconducting transition gradually shifts towards lower temperatures with increasing magnetic fields. At $\mu_0H = 3$ T, the transition could not be observed above 1.8 K. The upper critical field, $H_{c2}$, is determined using the 90% points on the resistive transition curves. A plot of $H_{c2}(T)$ is given in Supplementary Figure 3. The initial slope $d\mu_0H_{c2}/dT$ at $T_c$ is -1.08 T / K. A simple estimate using the conventional one-band Werthamer–Helfand–Hohenberg (WHH) approximation without considering the Pauli spin-paramagnetism effect and spin-orbit interaction[31], $H_{c2}(0) = -0.693T_c \times (dH_{c2}/dT)$ with $\mu_0H$ in Tesla and $T$ in Kelvin, yielded a value of 3.6 T. We also tried to use the Ginzburg-Landau formula to fit the data,

$$H_{c2}(T) = H_{c2}(0)\frac{1-t^2}{1+t^2}$$

where $t$ is the reduced temperature $T/T_c$. The resulting upper critical field $\mu_0H_{c2}(0) = 4.5$ T. These $H_{c2}$ values are obviously higher than that obtained in the sister compound $ZrTe_5$[19]. According to the relationship between $H_{c2}$ and the Ginzburg-Landau coherence length $\xi_{GL}$, namely, $H_{c2} = \Phi_0/(2\pi\xi^2)$, where $\Phi_0 = 2.07\times10^{-15}$ Wb is the flux quantum, the derived $\xi_{GL}(0)$ is 8.5 nm. It is also worth noting that our estimated value of $H_{c2}(0)$ is well below the Pauli-Clogston limit.

**Raman spectroscopy and possible crystal structures.** The described above changes of the electronic properties of $HfTe_5$ at high pressures might be associated with pressure-induced structural transitions. Raman spectroscopy is a powerful tool to probe changes in the crystal lattice and thus, our pressure-dependent electronic transport measurements of $HfTe_5$ were accompanied such spectroscopic studies. Figure 3a shows the Raman spectra of $HfTe_5$ at various pressures. The modes observed at the lowest experimental pressure of 0.5 GPa are similar to those reported previously at ambient pressure[32, 33]. At $P$ increase, the profile of the spectra remains similar to that at ambient pressure, whereas the observed modes shift toward higher frequencies, thus showing the normal pressure behavior. When the pressure approaches 4-5 GPa, the splitting of observed vibrational modes indicates the structural phase transition to high-pressure phase II. It should be noted that the superconductivity is observed beyond this pressure. An abrupt disappearance of Raman peaks for $P > 9$ GPa indicates the next structural phase transition to phase III. The absence of Raman peaks is consistent with the normal metallic state observed in our resistance measurements in this high-pressure phase above 9 GPa. To sum up, the Raman study provides evidence for two pressure-induced structural phase transitions.

Similarly, two structural phase transitions, from *Cmcm* to *C*2/*m* and to *P*-1, have been reported in recent high-pressure studies of ZrTe$_5$[19]. Considering the close similarities between ambient-pressure structure and high-pressure behavior of the electronic properties of the two compounds, it is natural to suppose that HfTe$_5$ adopts at high pressure the same crystal structures as ZrTe$_5$. Density functional theory (DFT) calculations of phase stabilities of HfTe$_5$ in these structures at high pressure confirm our suggestion. The enthalpy difference curves for the three phase are shown in Supplementary Figure 4. The enthalpy, *H*, of a given phase is evaluated to identify the energetically favored ground state for a finite pressure by $H = E_{tot} + PV$, where $E_{tot}$ is the total energy of the system and *V* the volume of a unit cell. The enthalpy-pressure curves indicate that the *Cmcm* structure is indeed the most stable one at the ambient condition, which agrees well with the experiment. In the pressure range from about 2.2 to 14.8 GPa, the orthorhombic *C*2/*m* phase has the lowest enthalpy but for higher pressure the *P*-1 phase takes over the ground state. The corresponding transition pressures are in good agreement with our Raman spectroscopy results. Since the *Cmcm* structure is experimentally verified for ambient pressure, we further verified the stabilities of the other two high-pressure phases by the phonon spectrum calculations (see Supplementary Figure 5) confirming that both *C*2/*m* and *P*-1 phases are dynamically stable. The crystal structures of *C*2/*m* and *P*-1 are shown in Figure 3b, c. The *Cmcm* and *C*2/*m* phases are similar in structure, in which a distorted square lattice of Te atoms exists. The Te layer is strongly corrugated in the *Cmcm* phase while it is relatively flat in the *C*2/*m* phase. By contrast, corresponding Te layer turns into one-dimensional chain-like structure in the *P*-1 phase with the lowest symmetry (Figure 3d, e).

**Pressure-temperature phase diagram.** The high-pressure experiments have been repeated on different samples with good reproducibility of observed transition temperatures. All the characteristic temperatures from our experiments are summarized in the *T-P* phase diagram in Figure 4. According to Raman spectroscopic data, there are two high-pressure phases (*C*2/*m*, phase II and *P*-1, phase III) in addition to the ambient-pressure phase (*Cmcm*, phase I). With increasing pressure the peak temperature of the resistivity initially increases to around 110 K and then decreases abruptly in phase II. Superconductivity appears with phase II, while the resistivity anomaly is still present a higher temperature. In phase III a metallic normal state is reached. The superconducting $T_c$ changes slowly with a maximum critical temperature of 4.8 K at $P \approx 20$ GPa and persists up to the highest pressure of 42 GPa with $T_c = 4.5$ K.

**Discussions**

The electronic band structure and density of states (DOS) can help to further understand the properties of $HfTe_5$. As shown in Supplementary Figure 6a, b, the *Cmcm* phase is semimetallic, in agreement with our resistivity and specific heat results. Here Hf-5*d* and Te-5*p* states exhibit a band anti-crossing near the Fermi energy $E_F$, which is consistent with the observed band inversion in a previous calculation[16]. However, the other two high-pressure phases are metallic and display large DOS at the Fermi energy $E_F$ (see Supplementary Figure 6c-f). Herein the states at $E_F$ are mainly contributed by the Te-5*p* states with negligible contribution of Hf-5*d* states. In the *C*2/*m* phase, the in-layer Te-5*p* states are dominant in the DOS at $E_F$ compared to those of the rest Te atoms. Similarly, in the *P*-1 phase the in-chain Te-5*p* states are dominant (see Supplementary Figure 6 g, h). We note that the abrupt increase of DOS at the transition points shown in Supplementary Figure 7 is due to that we only simulated the pressure behavior of neighboring phases of the transition, rather than the real continuous structural deformation from one phase to the other in experiment. However, we can still estimate the general trend of DOS: it increases up to the pressure region when the II-III transition happens and then decreases slowly, which agrees roughly with the $T_c$ in our superconductivity phase diagram. With increasing pressure the DOS increases suddenly when $HfTe_5$ transforms into the intermediate phase II structure. At this pressure superconductivity rises suddenly above our experimental low temperature limit. At the second transition (II-III), the DOS and $T_c$ further increase simultaneously. In phase III, DOS and $T_c$ decrease simultaneously with increasing pressure. Considering that the superconductivity occurs among the electronic states at the Fermi energy, the superconductivity of the $HfTe_5$ high-pressure phases may be hosted in different channels: inside the Te layers for the *C*2/*m* phase and among the Te chains for the *P*-1 phase. This connection of superconductivity to specific structural subunits of HfTe$_5$ resembles the situation in the 2D cuprates (CuO$_2$ planes)[34] and iron-pnictides (Fe$_2$As$_2$ layers)[35] on the one hand, and in some quasi-1D organic superconductors (as (TMTSF)$_2$X)[36] on the other hand. It may lead to interesting quasi-2D and quasi-1D superconducting properties of HfTe$_5$ under pressure.

In summary, metallicity and superconductivity was successfully induced in the semimetal HfTe$_5$ by application of high pressure. The appearance of superconductivity is accompanied by the suppression of the resistivity anomaly as well as by a structural phase transition. Thus, the resistivity anomaly, nontrivial topological state and superconductivity were all observed in HfTe$_5$, all contributing to the highly interesting physics seen in this transition-metal pentatelluride.

**Methods**

**Single–crystal growth and characterization.** Single crystals of HfTe$_5$ were prepared by a flux-growth method using Te as self-flux. In a typical synthesis, pieces of Hf and Te were weighed in a ratio Hf$_{0.0025}$Te$_{99.9975}$ and transferred to an alumina crucible inside an argon filled glove box. The crucible was then sealed inside a quartz tube under vacuum. The mixture was heated first to 900 °C for a day followed by rapid cooling to 580 °C. At this temperature further slow cooling with a rate of 0.5 K h$^{-1}$ was employed until 470 °C, where the excess of Te was decanted. Long ribbon shaped crystals of HfTe$_5$ were obtained. Elemental compositions were determined using energy-dispersive X-ray spectroscopy. The micrometer-scale compositions within the main phase were probed at 5-10 spots and the results were averaged. The structures of the HfTe$_5$ crystals were investigated using single-crystal X-ray diffraction with Mo $K_\alpha$ radiation. The electrical resistivity $\rho$ was measured using a four-probe method (low-frequency alternating current, PPMS, Quantum Design), the heat capacity was measured by a relaxation method (PPMS, Quantum Design).

**Experimental details of high-pressure measurements.** A non-magnetic diamond anvil cell was used for $\rho$ measurements under $P$ values of up to 42 GPa. A cubic BN/epoxy mixture was used for the insulating gaskets and Pt foil was employed for the electrical leads. The diameters of the flat working surface of the diamond anvil and the gasket were 0.5 and 0.2 mm, respectively, the sample chamber thickness was ≈ 0.04 mm. The value of $\rho$ was measured using direct current in van der Pauw technique in a customary cryogenic setup at zero magnetic field, and the magnetic-field measurements were performed on a PPMS. Pressure was measured using the ruby scale, for small chips of ruby placed in contact with the sample[37]. The high-$P$ Raman spectra were recorded using a customary micro-Raman spectrometer with a HeNe laser as the excitation source and a single-grating spectrograph with 1 cm$^{-1}$ resolution.

**Density-functional theory calculations.** Density-functional theory (DFT) calculations were performed using the Vienna Ab-initio Simulation Package (VASP) with plane wave basis[38]. The interactions between the valence electrons and ion cores were described by the projector augmented wave method[39, 40]. The exchange and correlation energy was formulated by the generalized gradient approximation with the Perdew-Burke-Ernzerhof scheme[41]. Van der Waals corrections were also included via a pair-wise force field of the Grimme method[42, 43]. The plane-wave basis cutoff energy was set to 283.0 eV. The Γ-centered $k$ points were used for the first Brillouin-zone sampling with a spacing of 0.03 Å$^{-1}$. The structures were optimized until the forces on atoms were less than 5 meV Å$^{-1}$. The pressure was derived by fitting the total energy dependence on the volume with the Murnaghan equation[44]. The phonon dispersion was carried out using the finite displacement method with VASP and PHOHOPY code[45], and a supercell with all the lattice constant larger than 10.0 Å was employed to calculate the phonon spectra.

**Acknowledgments**

Y. Qi acknowledge the financial support from the Alexander von Humboldt Foundation. This work was financially supported by the Deutsche Forschungsgemeinschaft DFG (Project No. EB 518/1-1 of DFG-SPP 1666 'Topological Insulators') and by the ERC Advanced Grant No. (291472) 'Idea Heusler'.

**Figure Captions**

**Figure 1. Crystal structure and resistivity of HfTe$_5$ at ambient pressure.** (**a**) Crystal structure of HfTe$_5$ with *Cmcm* space group. The red spheres represent Hf atoms and both green and blue spheres represent Te atoms at different crystallographic positions. (**b**) Side view of the HfTe$_5$ crystal structure. The HfTe$_3$ chains which run along the *a* axis and linked via zigzag chains of Te atoms. (**c**) temperature-dependent resistivity of HfTe$_5$ along the *a* axis. A large resistivity anomaly appears at around 40 K.

**Figure 2. Evolution of superconductivity as a function of pressure.** (**a**) The plot of electrical resistivity as a function of temperature. Inset: the resistivity curves at *P* = 5.5 GPa and 6.2 GPa. At this pressure, superconductivity is observed although the normal state still exhibits a pronounced resistivity anomaly. (**b**) shows the clear electrical resistivity drop and zero-resistivity behavior at low temperature. $T_c$ increases under increasing pressure and the maximum superconducting transition temperature $T_c$ = 4.8 K is observed at around 20 GPa. (**c**) Temperature dependence of the resistivity under different fields up to 4 T at 19.5 GPa.

**Figure 3. Raman spectroscopy of HfTe$_5$ and possible crystal structure under high pressure.** (**a**) Pressure-dependent Raman spectroscopic signals for HfTe$_5$ at room temperature. (**b**) and (**c**) show the crystal structure of the *C*2/*m* and *P*-1 phase. The red spheres represent Hf atoms and both green and blue spheres represent Te atoms with different positions. (**d**) and (**e**) show the two-dimensional Te-layers and the one-dimensional Te-chains.

**Figure 4. Electronic phase diagram of HfTe$_5$.** The black and magenta squares denote $T^*$, the peak temperature of electrical resistivity anomaly. The green and blue circles represent $T_c$ extracted from different runs of electrical resistivity measurements. Colored areas are a guide to the eye indicting the distinct phases.

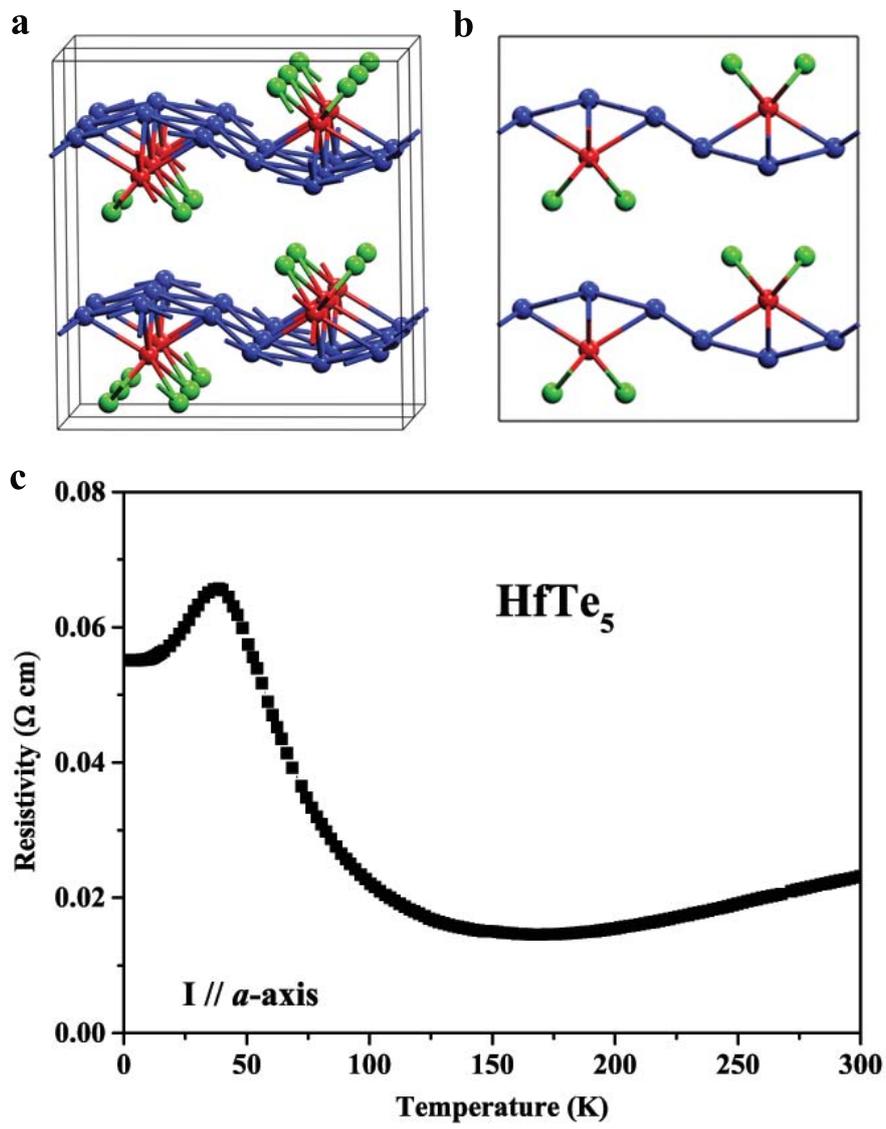

Fig. 1 Qi et al.

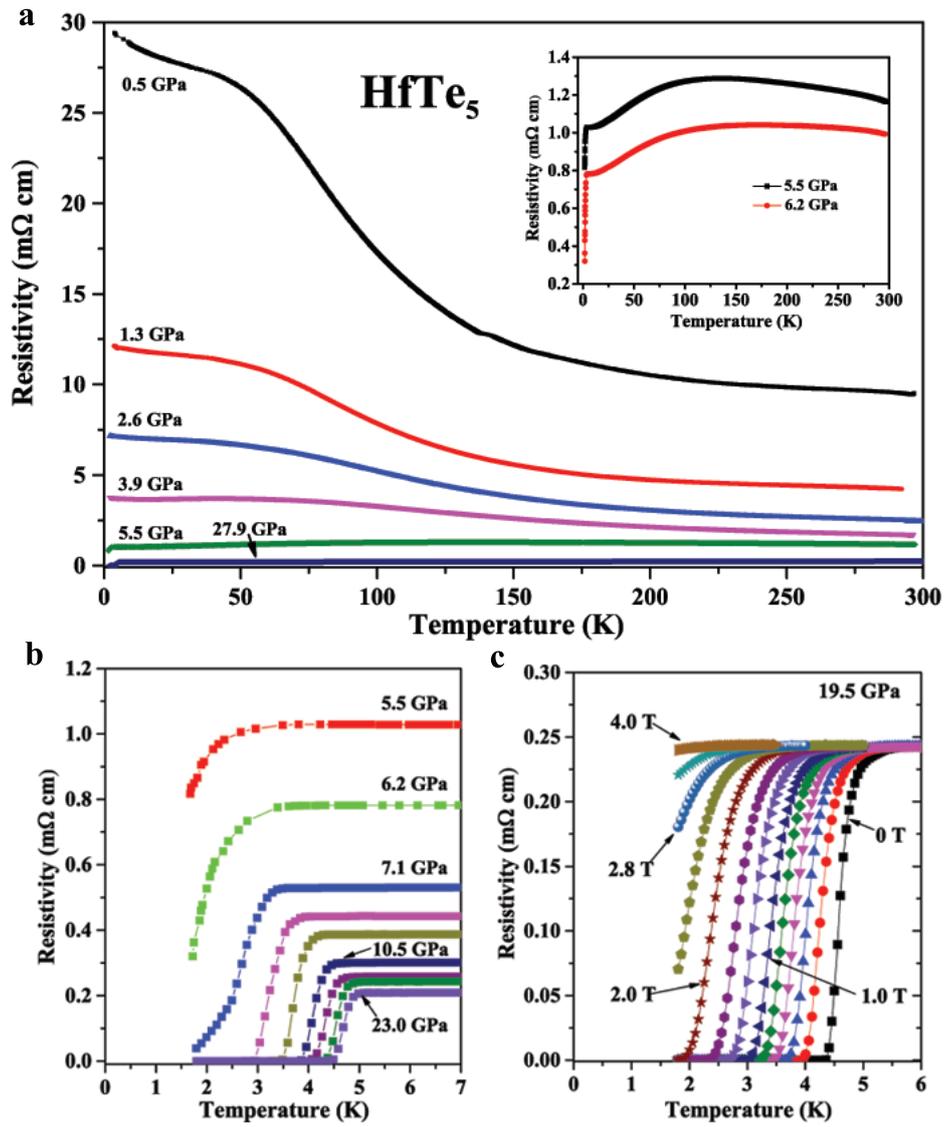

Fig. 2 Qi et al.

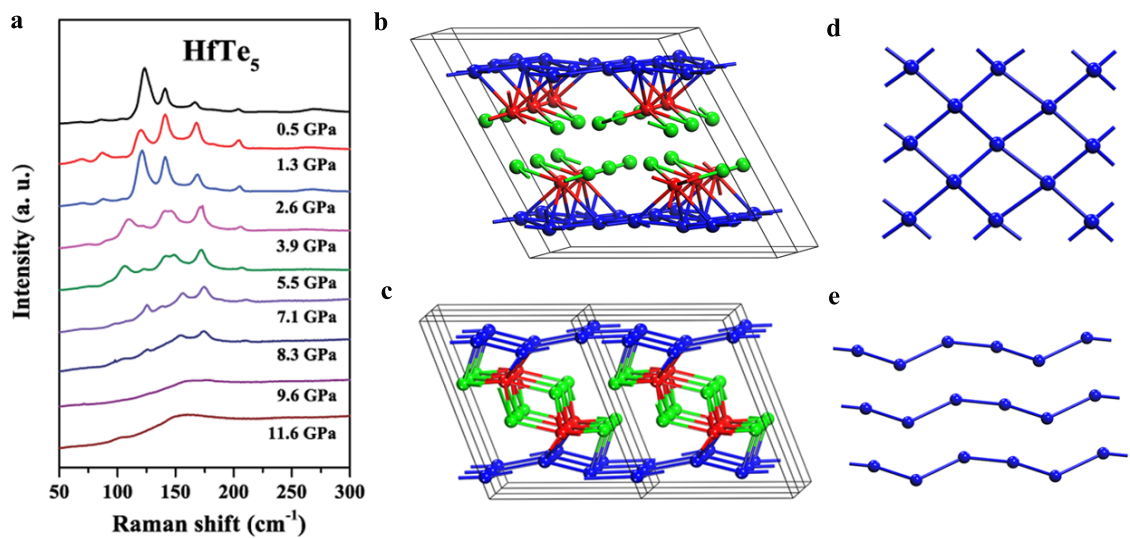

Fig. 3 Qi et al

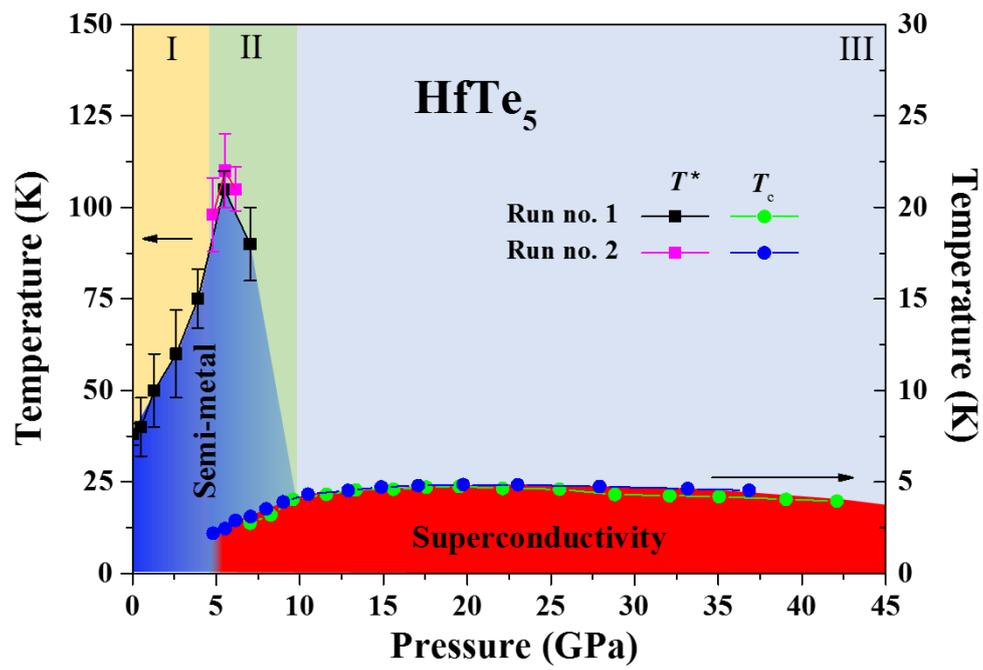

Fig. 4 Qi et al.